\documentclass[12pt]{article}
\usepackage{graphicx}
\usepackage{amsmath, amssymb}

\newcommand{\del}{\partial}

\DeclareMathOperator*{\Tr}{{\rm Tr}}
\DeclareMathOperator*{\tr}{{\rm tr}}

\newcommand{\vevmm}[1]{\left\langle#1\right\rangle_{mm}}
\newcommand{\Fcal}{{\cal F}}
\newcommand{\Jt}{\widetilde{J}}
\newcommand{\St}{\widetilde{S}}

\numberwithin{equation}{section}

\begin{document}

\thispagestyle{empty}

\begin{flushright}
 \begin{tabular}{l}
 {\tt hep-th/0701052}\\
 IHES/P/07/03
 \end{tabular}
\end{flushright}

 \vfill
 \begin{center}
{\LARGE
\begin{center}
  Semi-classical open string corrections \\ and symmetric Wilson loops
\end{center}
 \vskip 3mm
 \centerline{ }
}

 \vskip 2.0 truecm
\noindent{ \large  Satoshi Yamaguchi} \\
{\sf yamaguch@ihes.fr}
\bigskip

 \vskip .6 truecm
 {
 {\it 
IHES, Le Bois-Marie, 35, route de Chartres\\ 
F-91440 Bures-sur-Yvette,
FRANCE
} 
 }
 \vskip .4 truecm

 \end{center}

 \vfill
\vskip 0.5 truecm

\begin{abstract}
In the AdS/CFT correspondence, an $AdS_2\times S^2$ D3-brane with electric flux in $AdS_5\times S^5$ spacetime corresponds to a circular Wilson loop in the symmetric representation or a multiply wound one in $N=4$ super Yang-Mills theory. In order to distinguish the symmetric loop and the multiply wound loop, one should see an exponentially small correction in large 't Hooft coupling. We study semi-classically the disk open string attached to the D3-brane. We obtain the exponent of the term and it agrees with the result of the matrix model calculation of the symmetric Wilson loop.
\end{abstract}
\vfill
\vskip 0.5 truecm

\newpage

\section{Introduction and summary}

Wilson loops are some of the most interesting non-local operators in Yang-Mills theories. In particular the Wilson loops in $N=4$ super Yang-Mills theory have interesting string theory counterparts in the AdS/CFT correspondence \cite{Maldacena:1997re}. The Wilson loop in the fundamental representation corresponds to a macroscopic fundamental string in $AdS_5\times S^5$ spacetime \cite{Rey:1998ik,Maldacena:1998im}. Moreover the Wilson loops in higher rank representations were recently explored in the AdS/CFT correspondence, the D-brane probe descriptions \cite{Rey:1998ik,Drukker:2005kx,Hartnoll:2006hr,Yamaguchi:2006tq,Gomis:2006sb,Rodriguez-Gomez:2006zz,Okuyama:2006jc,Hartnoll:2006is,Hartnoll:2006ib,Chen:2006iu,Giombi:2006de,Tai:2006bt,Gomis:2006im,Drukker:2006zk,Tai:2007gg} and the supergravity descriptions \cite{Yamaguchi:2006te,Lunin:2006xr} were developed.

One of these D-brane probe pictures is the D3-brane description of the symmetric or multiply wound Wilson loop\cite{Drukker:2005kx}. These two different kind of Wilson loop (the multiply wound Wilson loop and symmetric one) have the same leading term in the expectation value in large $\lambda$ ('t~Hooft coupling) limit. In other words, the difference between these two VEVs is exponentially smaller, like $\exp[-\sqrt{\lambda}(\text{constant})]$, than the expectation values. It was shown in \cite{Gomis:2006im} that the D3-brane corresponds to the symmetric Wilson loop from the point of view of the low energy theory on the D-branes. In this letter we want to see this exponentially smaller term in the AdS side of the correspondence.

In order to treat this kind of small correction, one should include some quantum effects in the AdS side of the calculation. In principle, if one sums up all the configurations which satisfy a certain boundary condition at the AdS boundary, one gets the exact result. This set of the configurations includes small massless fluctuation, open strings, D-branes, new geometries and perhaps other things. They may or may not be a solution of the classical equation of motion. The only constraint is the boundary condition at the AdS boundary; the configuration becomes, in the problem here, the single D3-brane with electric flux at the boundary of the AdS. We consider the configuration with a disk open string worldsheet attached to the D3-brane among these configurations since $\exp[-\sqrt{\lambda}(\text{constant})]$ type corrections usually appear as worldsheet non-perturbative\footnote{Here ``worldsheet non-perturbative'' means non-perturbative in the worldsheet sense i.e. the correction like $\exp[-(\text{constant})/\alpha']$. It is not non-perturbative in the string theory sense.} corrections.

This kind of worldsheet non-perturbative effects typically appears as worldsheet instantons. In the Wilson loop literature, for example, the open string instanton in the 1/4 BPS Wilson loop captures the exponentially smaller term in the asymptotic expansion of the modified Bessel function \cite{Drukker:2006ga}. The worldsheet instantons also play a central role in the Wilson loops in the topological large $N$ duality \cite{Gomis:2006mv}.

In this letter, we find that there is also a worldsheet non-perturbative effect in the open string attached to the D3-brane which corresponds to the symmetric Wilson loop. It captures the exponentially small difference between the symmetric and multiply wound Wilson loops. Actually this difference is calculated using the Gaussian matrix model as eq. \eqref{matrix-result}. The open string non-perturbative effect is shown in eq. \eqref{disk-result}. Both of them are written as
\begin{align}
 \exp\left[-\sqrt{\lambda}\left(\sqrt{1+\kappa^2}-1\right)\right],
\end{align}
where $\kappa$ is defined as $\kappa:={k\sqrt{\lambda}}/(4N)$ and $k$ denotes the rank of the symmetric representation. This result support the statement that the D3-brane corresponds to the symmetric Wilson loop, but not the multiply wound Wilson loop.

The construction of this letter is as follows. In section \ref{sec-matrix} we use the Gaussian matrix model to evaluate the correction in the Yang-Mills theory side. In section \ref{sec-worldsheet}, we consider the open string on the D3-brane and evaluate the worldsheet non-perturbative correction.

\section{Matrix model calculation}
\label{sec-matrix}
In this section, we evaluate the leading difference of the rank $k$ symmetric Wilson loop and $k$-times wound Wilson loop in $N=4$ super Yang-Mills theory, by using the Gaussian matrix model\cite{Erickson:2000af,Drukker:2000rr}. We evaluate it in the following limit.
\begin{itemize}
 \item First, take $N\to \infty$ with $\lambda:=g_{YM}^2N$ and $k/N$ kept finite.
 \item Then, take $\lambda\to \infty$ with $\kappa:=k\sqrt{\lambda}/(4N)$ kept finite.
\end{itemize}
This quantity can be calculated in various method (see for example \cite{Drukker:2000rr,Okuyama:2006jc,Hartnoll:2006is} and references therein). Here we put emphasis on the eigenvalue distribution in order to see the pictorial similarity to the AdS side in the next section.

We consider the 1/2 BPS circular Wilson loop $\Tr_{R} U$ with the representation $R$ of $SU(N)$ and the group element $U$ is defined as
\begin{align}
 U:=P\exp \int_{C} d\tau (\dot{x}^{\mu}(\tau)iA_{\mu}(x(\tau))+|\dot{x}(\tau)|\phi_1(x(\tau))),
\end{align}
where the trajectory $C$ is a circle, $\dot{x}$ denotes the differential $dx^{\mu}/d\tau$, $A_{\mu}$ is the gauge field and $\phi_1$ is one of the six scalar fields in $N=4$ super Yang-Mills theory.

Let us introduce some notations in order to express the gauge invariant polynomials. Let $u_1,u_2,\dots,u_N$ be the eigenvalues of the matrix $U$. Then for the choice of integers $\mu=(\mu_1,\mu_2,\dots,\mu_r),\ r\le N,\ 
\mu_1\ge \mu_2 \ge \dots \ge \mu_r$ the symmetric monomial $m_{\mu}(U)$ is defined as
\begin{align}
 m_{\mu}(U)=u_1^{\mu_1}u_2^{\mu_2}\dots u_r^{\mu_r} + (\text{symmetrization}).
\end{align}
The k-times wound Wilson loop is expressed by these notations as $\tr U^k=m_{(k)}(U)=\sum_{i=1}^{N}u_i^k$. On the other hand, the rank $k$ symmetric Wilson loop $\Tr_{S_k} U$ is the sum of all the monomials of degree $k$ as
\begin{align}
 \Tr_{S_k} U
=\sum_{\mu: \ k\text{ boxes}} m_{\mu}(U)
=m_{(k)}(U)+m_{(k-1,1)}(U)+\cdots.
\end{align}
Therefore in order to distinguish $\Tr_{S_k} U$ and $m_{(k)}(U)$ we should see $\left\langle m_{(k-1,1)}(U)\right\rangle/\left\langle m_{(k)}(U)\right\rangle$. This is what we want to evaluate here.

It is conjectured\cite{Erickson:2000af,Drukker:2000rr} that the expectation value of this Wilson loop is calculated by the Gaussian matrix model. Let $Y$ be an $N\times N$ Hermitian matrix and ``$\tr$'' be the trace in fundamental representation. The expectation value is calculated as
\begin{align}
 &\left\langle m_{\mu}(U) \right\rangle=\vevmm{m_{\mu}(e^{Y})}
:=\frac{1}{Z}\int dY \; m_{\mu}(e^{Y})
     \exp\left(-\frac{2N}{\lambda}\tr [Y^2]\right),\\
&Z:=\int dY \exp\left(-\frac{2N}{\lambda}\tr [Y^2]\right).
\end{align}
The standard method to evaluate this integral is to diagonalize the matrix $Y$.
The matrix integral above is rewritten in terms of the eigenvalues
\begin{align}
 &\vevmm{m_{\mu}(e^{Y})}=\frac{1}{Z}\int \prod_{i=1}^N dy_i \; m_{\mu}(e^{y})
     \exp\left(-I[y]\right),\\
&Z:=\int \prod_{i=1}^N dy_i 
     \exp\left(-I[y]\right),\\
&I[y]:=\frac{2N}{\lambda}\sum_{i=1}^{N}y_i^2-2\sum_{i<j}\log|y_i-y_j|.
\end{align}

\begin{figure}
 \begin{tabular}{cc}
 \includegraphics[width=6truecm]{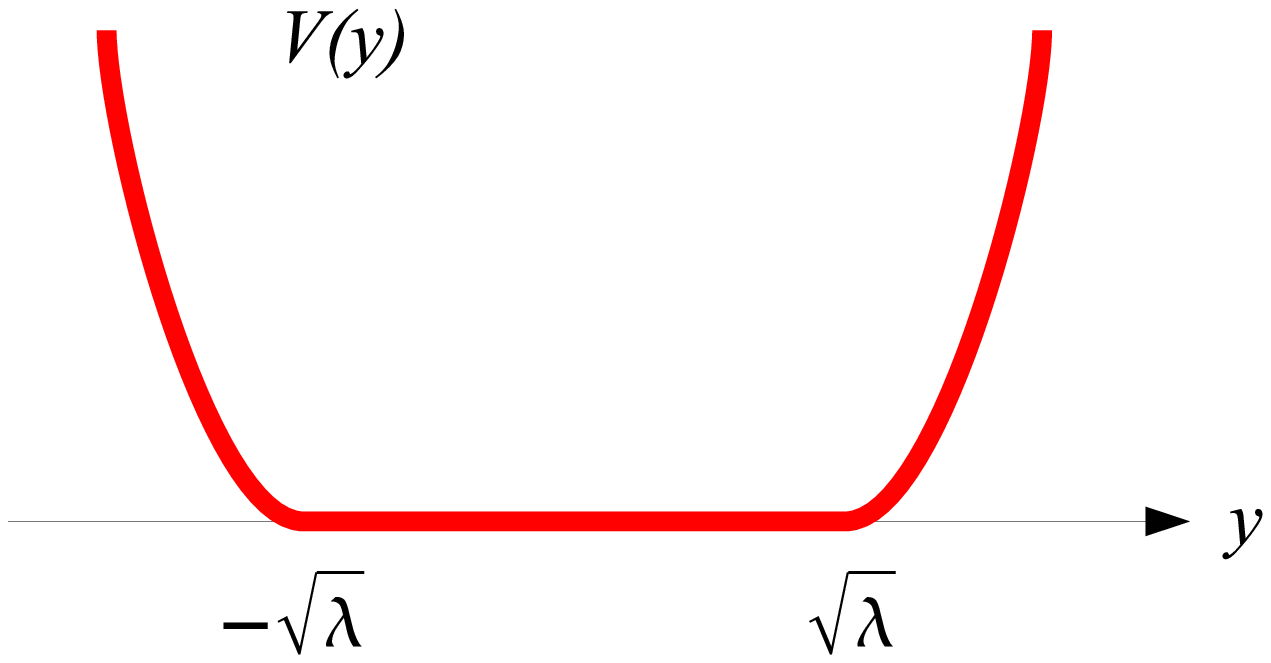} & \includegraphics[width=7truecm]{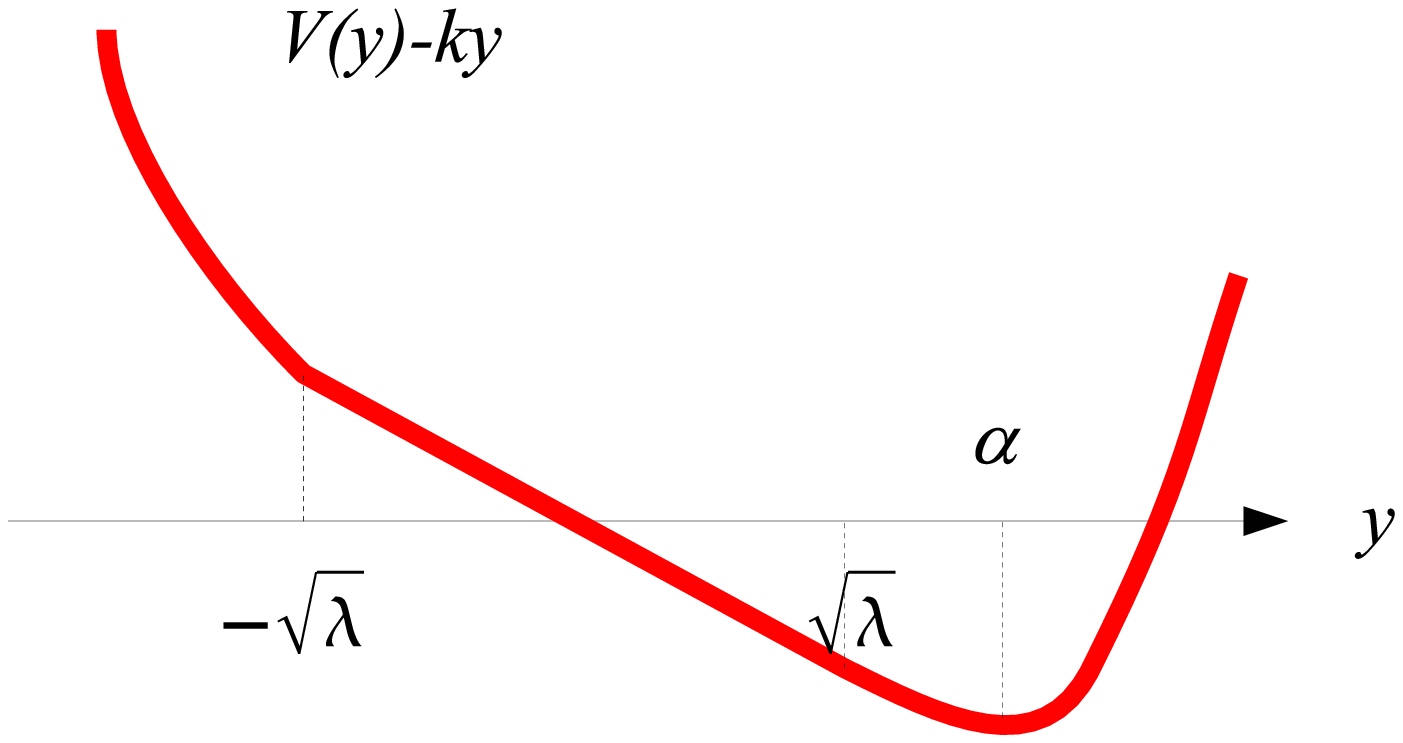}\\
 (a) & (b)
 \end{tabular}
\caption{(a) The graph of the effective potential $V(y)$. (b) The graph of $V(y)-ky$ when $k/N$ is finite.}
\label{vy}
\end{figure}

First we evaluate the $k$-times wound loop $\vevmm{m_{(k)}(e^Y)}$. 
\begin{align}
 \vevmm{m_{(k)}(e^Y)}
=\frac{N}{Z}\int \prod_{i=1}^N dy_i
     \exp\left(-I[y]+ky_1\right).\label{int1}
\end{align}
It is convenient to integrate out $y_2,\dots, y_N$ and consider the ``effective potential'' $V(y_1)$ for $y_1$ defined as
\begin{align}
\exp(-V(y_1))
:=\frac{N}{Z}\int \prod_{i=2}^N dy_i
     \exp\left(-I[y]\right).\label{veff}
\end{align}
Using this effective potential, the integral \eqref{int1} can be written as
\begin{align}
 \vevmm{m_{(k)}(e^Y)}
=\int_{-\infty}^{\infty} d y_1
\exp(-V(y_1)+ky_1).\label{y1int}
\end{align}
The behavior of $V(y)$ in the large $N$ are shown as figure \ref{vy} (a). In the region $-\sqrt{\lambda}\le y \le \sqrt{\lambda}$, the potential is almost flat compared to $N$. Actually, it becomes the Wigner's semi-circle distribution in the large $N$ limit
\begin{align}
 \exp(-V(y))=\rho(y),\qquad (-\sqrt{\lambda}\le y \le \sqrt{\lambda}),\\
 \rho(y):=\frac{2N}{\pi \lambda}\sqrt{\lambda-y^2}. 
\end{align}
 On the other hand, the potential $V(y_1)$ in $y_1\ge \sqrt{\lambda}$ is obtained by applying the saddle point approximation to the integral \eqref{veff}. It is expressed as
\begin{align}
 V(y_1)=\frac{2N}{\lambda}y_1^2-2\int_{-\sqrt{\lambda}}^{\sqrt{\lambda}} dz \rho(z)\log(y_1-z),\qquad (y_1\ge\sqrt{\lambda}).
\end{align}
This potential in this region is proportional to $N$ in the large $N$ limit.

If we consider the large $N$ limit with $k$ kept finite, $ky_1$ is much smaller than $N$, the wall of $V(y_1)$ shown figure \ref{vy} (a) becomes steep enough. Then the integral \ref{y1int} can be evaluated in the bottom of the potential as
\begin{align}
 \int_{-\infty}^{\infty} d y_1
\exp(-V(y_1)+ky_1)=\int_{-\sqrt{\lambda}}^{\sqrt{\lambda}} d y_1
\rho(y_1)\exp(ky_1).\label{smallk}
\end{align}
This integral is expressed by the modified Bessel function.

However, what we really want to do is to take the limit $N \to \infty,\ k\to \infty$ with $k/N$ kept finite. In this case, $V(y_1)$ and $ky_1$ are in the same order in $N$ (linear in $N$). Therefore we can not ignore the term $ky_1$ when searching the saddle point in large $N$. The function $V(y_1)-ky_1$ looks like figure \ref{vy} (b). This function is not flat at all in the region $-\sqrt{\lambda}\le y \le \sqrt{\lambda}$. Thus the expression \eqref{smallk} is not valid in this case. We find the minimum at a certain point $y_1=\alpha\ge \sqrt{\lambda}$ instead.

Let us evaluate the integral \eqref{y1int} by the point $y_1=\alpha$.
$\alpha$ is the solution of the equation
\begin{align}
 0=\frac{\del}{\del y_1}(V(y_1)-ky_1)=\frac{4N}{\lambda}y_1-2\int_{-\sqrt{\lambda}}^{\sqrt{\lambda}} dz\frac{\rho(z)}{y_1-z}-k=\frac{4N}{\lambda}\sqrt{y_1^2-\lambda}-k.
\label{eom1}
\end{align}
Then eq.\eqref{eom1} is solved as
\begin{align}
 y_1=\alpha(k):=\sqrt{\lambda}\sqrt{1+\kappa^2},\qquad \kappa=\frac{k\sqrt{\lambda}}{4N}.
\label{sol1}
\end{align}
Here we add the argument to $\alpha(k)$ in order to remember that it depends on $k$. As a result, we can approximate the integral \eqref{y1int} in the large $N$ limit as
\begin{align}
 \vevmm{m_{(k)}(e^Y)}=\exp[-V(\alpha(k))+k \alpha(k)]=\exp[2N(\kappa\sqrt{1+\kappa^2}+\sinh^{-1}\kappa)].
\label{mk}
\end{align}
This is the same result as in \cite{Drukker:2005kx}.
For later convenience, we define the quantity $\Fcal_{(k)}$ as
\begin{align}
 \Fcal_{(k)}=-2N(\kappa\sqrt{1+\kappa^2}+\sinh^{-1}\kappa).
\end{align}

Next, we turn to the calculation of $\vevmm{m_{(k-1,1)}(e^{Y})}$. This expectation value is written as
\begin{align}
 \vevmm{m_{(k-1,1)}(e^Y)}=\frac{N(N-1)}{Z}
 \int \prod_{i=1}^N dy_i
     \exp\left(-I[y]+(k-1)y_1+y_2\right).\label{int2}
\end{align}
As the same way as above, let us define the effective potential $V_2(y_1,y_2)$ by
\begin{align}
 \exp(-V_2(y_1,y_2))=\frac{N(N-1)}{Z}
 \int \prod_{i=3}^N dy_i
     \exp\left(-I[y]\right). \label{eff2}
\end{align}
Then eq.\eqref{int2} is written as
\begin{align}
 \vevmm{m_{(k-1,1)}(e^Y)}=\int_{-\infty}^{\infty}dy_1\int_{-\infty}^{\infty}dy_2
\exp(-V_2(y_1,y_2)+(k-1)y_1+y_2). \label{y1y2int}
\end{align}
Actually, in the large $N$ limit, eq. \eqref{eff2} leads to the following expression of $V_{2}(y_1,y_2)$.
\begin{align}
 V_{2}(y_1,y_2)=V(y_1)+V(y_2)-2\log|y_1-y_2|.
\end{align}
Next let us turn to the $y_1,y_2$ integral \eqref{y1y2int}. We perform the $y_1$ integral first by the saddle point $y_1=\alpha(k-1)$. As for $y_2$ integral, the approximation like eq. \eqref{smallk} is valid since the situation is similar to eq. \eqref{smallk} with $k=1$. As a result, eq. \eqref{y1y2int} becomes
\begin{align}
 \vevmm{m_{(k-1,1)}(e^Y)}=\exp(-\Fcal_{(k-1)})\int_{-\sqrt{\lambda}}^{\sqrt{\lambda}}dy_2 \rho(y_2)(\alpha(k-1)-y_2)^2\exp(y_2). \label{y2int}
\end{align}
So far we only take large $N$ limit with $k/N$ and $\lambda$ kept finite. Thus, the expression \eqref{y2int} is valid for finite $\lambda$. Now we take the large $\lambda$ limit with $\kappa=k\sqrt{\lambda}/4N$ kept finite. In this limit, we can use the saddle point approximation at $y_2=\sqrt{\lambda}$. The integral \eqref{y2int} become
\begin{align}
 \vevmm{m_{(k-1,1)}(e^Y)}=\exp\left[-\Fcal_{(k-1)}+\sqrt{\lambda}\right]
 =\exp\left[-\Fcal_{(k)}-\sqrt{\lambda}\sqrt{1+\kappa^2}+\sqrt{\lambda}\right].
\end{align}
The final result of this section is
\begin{align}
 \frac{\left\langle m_{(k-1,1)}(U)\right\rangle}{\left\langle m_{(k)}(U)\right\rangle}
= \frac{\vevmm{m_{(k-1,1)}(e^{Y})}}{\vevmm{m_{(k)}(e^{Y})}}
 =\exp\left[-\sqrt{\lambda}(\sqrt{1+\kappa^2}-1)\right].
\label{matrix-result}
\end{align}
This is actually exponentially small in the large $\lambda$ limit.
In the next section, we will compare this exponent to the worldsheet non-perturbative correction.

\begin{figure}
 \begin{center}
  \includegraphics[width=10cm]{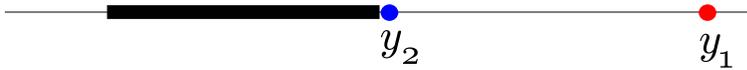}
 \end{center}
 \caption{The eigenvalue distribution which dominates the expectation value $\vevmm{m_{(k-1,1)}(e^{Y})}$. The black thick line denotes the semi-circle distribution of eigenvalues $y_3,\dots,y_N$.}\label{eigen-matrix}
\end{figure}

Let us make a comment on the eigenvalue distribution here. In the large $N$ and large $\lambda$ limit, the eigenvalue distribution at the saddle point looks like figure \ref{eigen-matrix}. This will be compared to the brane configuration in the AdS side of the calculation.
\section{Disk worldsheet corrections}
\label{sec-worldsheet}
In this section, we will consider the disk open string whose boundary is attached to the $AdS_2\times S^2$ D3-brane, and compute the disk open string amplitudes by using the semi-classical technique. We concentrate on the exponent of the correction in the large $\lambda$ limit. We postpone the integral on the moduli space and the one-loop determinant to future works. We find that the exponent of the correction agree with the matrix model result \eqref{matrix-result}.

Actually, there are perturbative corrections to the expression \eqref{mk}. We might not be allowed to retain the non-perturbative correction calculated in this section, since it is much smaller than the perturbative corrections. This may be justified by the supersymmetry but we leave it to future works.

It is convenient to use the coordinate system of \cite{Drukker:2005kx}. The metric of $AdS_5$ is expressed as
\begin{align}
\begin{aligned}
 ds^2=\frac{L^2}{\sin^2\eta}\left[
d\eta^2+\cos^2\eta d\psi^2+d\rho^2+\sinh^2\rho(d\theta^2+\sin^2\theta d\phi)\right],\qquad L^2=\alpha'\sqrt{\lambda},\\
0\le \eta \le \frac{\pi}{2},\qquad
0\le \psi \le 2\pi,\qquad
0\le \rho,\qquad
0\le \theta\le \pi,\qquad
0\le \phi \le 2\pi.
\end{aligned}
 \label{metric}
\end{align}
The $AdS_2\times S^2$ D3-brane worldvolume \cite{Rey:1998ik,Drukker:2005kx} is expressed in this coordinate system as
\begin{align}
 \sinh\rho=\kappa \sin \eta,\qquad \kappa=\frac{k\sqrt{\lambda}}{4N}.
\label{d3}
\end{align}
The electric field on the D3-brane worldvolume is excited and takes the value
\begin{align}
 F=dA=i\frac{\sqrt{\lambda}}{2\pi}\frac{\kappa}{\sinh^2\rho}d\psi d\rho,
 \label{electric}
\end{align}
where we identify $\psi$ and $\rho$ as worldvolume coordinates.

Now let us consider the disk string worldsheet whose boundary is attached to the D3-brane expressed by eqs.\eqref{d3},\eqref{electric}. Let the worldsheet coordinates be $(\sigma,\chi)$. The coordinate $\sigma,\ (\sigma_0\le \sigma \le \sigma_1)$ is the radial coordinate of the disk. $\sigma=\sigma_0$ corresponds to the boundary, while $\sigma=\sigma_1$ to the center of the disk. The other coordinate $\chi,\ (0\le \chi \le 2\pi)$ is the angular coordinate of the disk.
The string worldsheet action is written as
\begin{align}
 S=S_{bulk}+S_{bdy},\qquad
S_{bulk}=\frac{1}{2\pi\alpha'}\int d\sigma d\chi
\sqrt{\det G},\qquad
S_{bdy}=i \int_{\sigma=\sigma_0} A,\label{act1}
\end{align}
where $G$ is the induced metric and $A$ is the gauge field \eqref{electric}.

In this letter, we consider the following special ansatz.
\begin{align}
 \eta=\eta(\sigma),\qquad \rho=\rho(\sigma),\qquad
 \psi=\chi,\qquad \theta=0.\label{ansatz}
\end{align}
Since the center of the worldsheet $\sigma=\sigma_1$ is one point and should be mapped to one point in spacetime, the condition $\eta(\sigma_1)=\pi/2$ is imposed. Meanwhile the boundary value of $\eta$ is denoted by $\eta(\sigma_0)=\eta_0$. This boundary of the string is attached to the D3-brane \eqref{d3}, and it gives a constraint on the boundary value of
$\rho$ as $\rho(\sigma_0)=\sinh^{-1}(\kappa \sin \eta_0)$.

Putting this ansatz into the action \eqref{act1}, we obtain (prime ``\;${}'$\;'' denotes the $\sigma$ derivative)
\begin{align}
 &S_{bulk}=\sqrt{\lambda}\int_{\sigma_0}^{\sigma_1}d\sigma
\frac{\cos \eta}{\sin^2\eta}\sqrt{\eta'^2+\rho'^2},\label{act2bulk}
\\
 &S_{bdy}=\sqrt{\lambda}
 \left(\sqrt{1+\kappa^2}-\frac{\sqrt{1+\kappa^2\sin^2\eta_0}}{\sin\eta_0}\right).
\label{act2bdy}
\end{align}
The constant shift of the boundary action \eqref{act2bdy} is fixed so that
$S_{bdy}=0$ at $\eta_0=\pi/2$ where the boundary of the worldsheet shrinks to a point.

\begin{figure}
 \begin{center}
  \includegraphics[width=8cm]{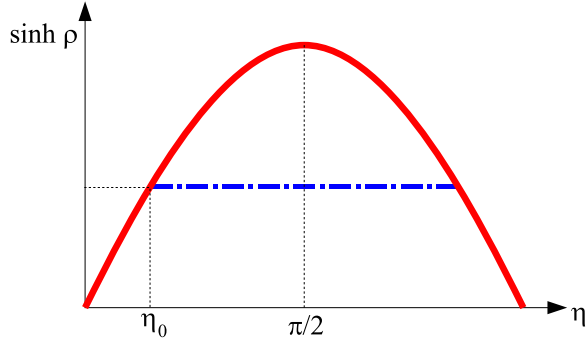}
 \end{center}
 \caption{The configuration \eqref{saturate} is shown in the $\rho$-$\eta$ plane. The red solid line expresses the D3-brane \eqref{d3}, while the blue dash-dotted line expresses the open string \eqref{saturate} ending on the D3-brane at $\eta=\eta_0$.}\label{fig-bubble}
\end{figure}

If we fix the boundary value $\eta_0$, the bulk action \eqref{act2bulk} has the lower bound
\begin{align}
 S_{bulk}\ge &\sqrt{\lambda}\int_{\sigma_0}^{\sigma_1}d\sigma
\frac{\cos \eta}{\sin^2\eta}\eta'
=\sqrt{\lambda}\left(-1+\frac{1}{\sin\eta_0}\right)\label{bound}.
\end{align}
This bound \eqref{bound} is saturated when
\begin{align}
 \rho'=0 \qquad \Rightarrow \qquad \rho=(\text{constant})=\sinh^{-1}(\kappa \sin \eta_0)
\label{saturate}.
\end{align}
This configuration \eqref{saturate} actually satisfies the equations of motion derived from the bulk action \eqref{act2bulk}. This configuration is shown in figure \ref{fig-bubble}.

When we derive the bound \eqref{bound}, we assume the boundary $\eta(\sigma_0)=\eta_0$ is fixed. However this is not the true boundary condition; the boundary of the string worldsheet can move along the D3-brane.
In this sense the configuration \eqref{saturate} is not a stationary point of the action.

Though the configuration \eqref{saturate} is not a solution, it is still useful to evaluate the path-integral; it is ``the bottom of the trough'' \cite{Coleman:1978ae}. We explain here how to evaluate the path-integral using the configuration \eqref{saturate}. We want to evaluate the path-integral
\begin{align}
 J=\int D\eta D\rho \;\exp\left(-S\right),
\end{align}
with the correct boundary condition determined by the configuration of the D3-brane. This integral can be rewritten as
\begin{align}
 J=\int d\eta_0 \Jt(\eta_0),\qquad
 \Jt(\eta_0)=\int_{\eta(\sigma_0)=\eta_0} D\eta D\rho \;\exp\left(-S\right).
\label{J}
\end{align}
In the path-integral $\Jt(\eta_0)$ the boundary value of $\eta$ is fixed to $\eta_0$. Hence \eqref{saturate} is the saddle point of this integral. The path-integral $\Jt(\eta_0)$ can be evaluated by the point. 
\begin{align}
 &\Jt(\eta_0)=\exp\left(-\St(\eta_0)\right),\\
 &\St(\eta_0)=(S_{bulk}+S_{bdy})|_{\rho=\sinh^{-1}(\kappa \sin \eta_0)}
=\sqrt{\lambda}\left(\sqrt{1+\kappa^2}-1
+\frac{1-\sqrt{1+\kappa^2\sin^2\eta_0}}{\sin\eta_0}\right).
\end{align}
The $\eta_0$ integral in \eqref{J} is written as
\begin{align}
 J=\int_{0}^{\pi/2}d\eta_0\exp[-\St(\eta_0)]=
\exp[-T(\pi/2)]-\exp[-T(0)],\label{T}
\end{align}
where $T(\eta_0)$ is defined as a solution of
\begin{align}
 T(\eta_0)-\log \frac{d T}{d \eta_0}(\eta_0)=\St(\eta_0).
\end{align}
Since $\St(\eta_0)$ is proportional to $\sqrt{\lambda}$, we can take $T(\eta_0)$ as
\begin{align}
 T(\eta_0)=\St(\eta_0)+O(\log \sqrt{\lambda}).
\end{align}
The first term on the right-hand side of eq.\eqref{T} is
\begin{align}
 \exp[-T(\pi/2)]=\exp[-\St(\pi/2)+O(\log \sqrt{\lambda})]
 =(\text{powers of }\lambda).
\end{align}
Thus this term captures the perturbative corrections. The second term on the right-hand side of eq.\eqref{T} is the exponentially small term that we want to see here. It is written as
\begin{align}
 \exp[-T(0)]=\exp\left[-\St(0)+O(\log \sqrt{\lambda})\right]
 \cong \exp\left[-\sqrt{\lambda}\left(\sqrt{1+\kappa^2}-1\right)\right].
 \label{disk-result}
\end{align}
This is the main result of this letter. The result \eqref{disk-result} agrees with the matrix model result \eqref{matrix-result}.

This small non-perturbative effect can be understood qualitatively as follows. There are two forces acting on the string end point: the string tension pulling the end point inside, and the electric force pushing the end point outside. The string tension is always larger than the electric force, and there is no stationary point other than the constant map. However, as the string worldsheet becomes larger and larger, the difference of these two forces becomes smaller and smaller. Actually when the worldsheet is large enough ($\eta_0 \ll 1$), the two forces almost cancel each other and the worldsheet boundary can be moved almost freely without increasing or decreasing the action. In other words, however large the worldsheet becomes, the action remains finite. The correction \eqref{disk-result} is the result of this effect.

This kind of exotic effect is not present for a flat D-brane in the flat space. In this case, the action diverges as the worldsheet becomes larger. Thus the contribution to the amplitudes is zero. Therefore the DBI action does not capture this effect since it is based on the small curvature approximation.

\begin{figure}[htb]
 \begin{center}
  \includegraphics[width=8cm]{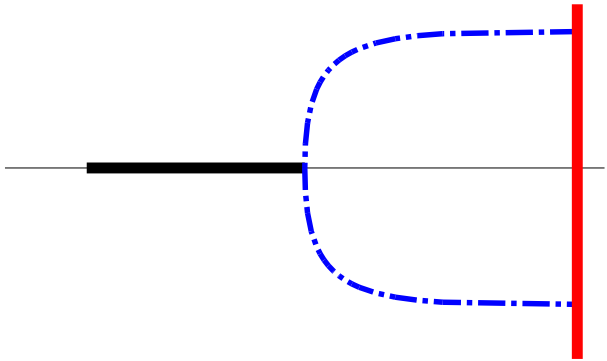}
 \end{center}
 \caption{The configuration \eqref{saturate} in the picture of \cite{Yamaguchi:2006te}. The horizontal direction is ``$x$'' of \cite{Yamaguchi:2006te}, while vertical direction is the radial direction of $AdS_2$ fiber. The D3-brane is represented by the red solid vertical line. The fundamental string is represented by the blue dash-dotted line.}\label{eigen-string}
\end{figure}

There is an intuitive explanation why the configuration \eqref{saturate} with $\eta_0\sim 0$ produces the term $\vevmm{m_{(k-1,1)}(e^{Y})}$. It is proposed in
\cite{Yamaguchi:2006te}\footnote{Taking the result of this letter into account, the black and white pattern in supergravity solution in \cite{Yamaguchi:2006te} interpreted as the eigenvalue distribution of the leading monomial of the representation expressed by a Young diagram. We can guess that there are corrections from the large closed strings, D-branes, geometry and so on. This is also an interesting future problem.} how the eigenvalue distribution of the Gaussian matrix model can be seen in $AdS_5\times S^5$. Figure \ref{eigen-string} represents the configuration \eqref{saturate} in the picture of \cite{Yamaguchi:2006te}. When $\eta_0$ is close enough to $0$, this picture around center is similar to figure \ref{eigen-matrix}; the fundamental string looks like $y_2$ and the D3-brane look like $y_1$ around the center\footnote{This fundamental string, even when $\eta_0\to 0$, is supposed to be different from the one used in the circular fundamental Wilson loop \cite{Berenstein:1998ij}. One can distinguish these two, for example, by the value of the action. Our string attached to the D3-brane has the action $\lim_{\eta_0\to 0} \St(\eta_0)=\sqrt{\lambda}\left(\sqrt{1+\kappa^2}-1\right)$. On the other hand the on-shell action of the string which represent the fundamental Wilson loop is $(-\sqrt{\lambda})$.}.

\begin{figure}[htb]
 \begin{center}
  \includegraphics[width=8cm]{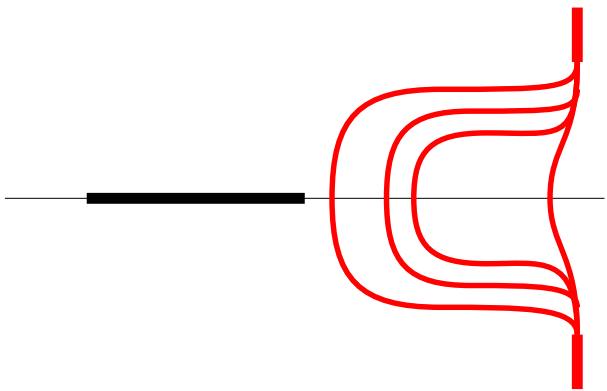}
 \end{center}
 \caption{The branched D3-brane configuration expected to correspond to a monomial in the symmetric representation.}\label{branched-d3}
\end{figure}

From the similar point of view, one may guess a branched D3-brane configuration, shown in figure \ref{branched-d3}, contributes as another monomial in the symmetric Wilson loop. Calculating this contribution is an interesting future problem.

\subsection*{Acknowledgment}
I would like to thank Gordon W. Semenoff for useful discussions. I am also grateful to Nadav Drukker for careful reading of the manuscript and useful comments.
This work was supported in part by the European Research Training Network contract 005104 ``ForcesUniverse.''

\providecommand{\href}[2]{#2}\begingroup\raggedright\endgroup

\end{document}